\documentclass[letterpaper]{ptephy_v1}
\usepackage{boites}
\usepackage{graphicx}
\usepackage{bm}
\usepackage{amsmath,amssymb,amsfonts}

\newcommand{\nc}{\newcommand}		% new command
\nc{\vc}[1]	{\mbox{\boldmath $#1$}}	% boldmath(vector)
\nc{\bra}       {\langle}               % bra state
\nc{\ket}       {\rangle}               % ket state
\nc{\del}       {\partial}              % bra state
\nc{\AMD}       {{\rm AMD}}
\nc{\TOAMD}     {{\rm TOAMD}}
\newcommand{\lw}[1]{\smash{\lower1.75ex\hbox{#1}}}

\usepackage{color}  % for color  usage: \textcolor{red}{text}
\begin{document}

\title{
Short-range correlation in high-momentum antisymmetrized molecular dynamics
}

\author{\name{Takayuki Myo}{1,2}
}
%%%%%%%%%%% The \name command should be used as \name{Insert author name here}{Insert affiliation number here}
%%%%% Please use \thanks for contributed author details

%%%%%%%%%%% The \affil command should be used as \affil{Insert affiliation number here}{Insert author address here}
\address{\affil{1}{General Education, Faculty of Engineering, Osaka Institute of Technology, Osaka, Osaka 535-8585, Japan}
\\
\affil{2}{Research Center for Nuclear Physics (RCNP), Osaka University, Ibaraki, Osaka 567-0047, Japan}
\email{takayuki.myo@oit.ac.jp}}

\begin{abstract}%
We propose a new variational method for treating short-range repulsion of bare nuclear force for nuclei in antisymmetrized molecular dynamics (AMD).
In AMD, the short-range correlation is described in terms of large imaginary centroids of Gaussian wave packets of nucleon pairs in opposite signs, causing high-momentum components in nucleon pair.
We superpose these AMD basis states and name this method ``high-momentum AMD'' (HM-AMD),
which is capable of describing strong tensor correlation (Prog. Theor. Exp. Phys., {\bf 2017}, 111D01 (2017)).
In this paper, we extend HM-AMD by including up to two kinds of nucleon pairs in each AMD basis state
utilizing the cluster expansion, which produces many-body correlations involving high-momentum components.
We investigate how much HM-AMD describes the short-range correlation by showing the results for $^3$H using the Argonne V4$^\prime$ central potential.
It is found that HM-AMD reproduces the results of few-body calculations and also the tensor-optimized AMD.
This means that HM-AMD is a powerful approach to describe the short-range correlation in nuclei. 
In HM-AMD, momentum directions of nucleon pairs isotropically contribute to the short-range correlation,
which is different from the tensor correlation.
\end{abstract}

\subjectindex{ D10,  D11}
% D10  Nuclear many-body theories
% D11  Models of nuclear structure

%\parindent0pt

\maketitle
%%%%%%%%%%%%%%%%%%%%%%%%%%%%%
\section{Introduction}
Bare nucleon-nucleon ($NN$) interaction has a strong short-range repulsion and a strong tensor force \cite{pieper01}.
Short-range repulsion causes short-range correlation in nuclei by reducing the short-range amplitudes in nucleon pair. 
Tensor force causes a spatially compact $D$-state via a strong $S$-$D$ coupling as tensor correlation.
These two correlations commonly induce the high-momentum components in nuclei \cite{pieper01_2}.

Recently, we developed two kinds of new variational methods for nuclei \cite{myo15,myo17a,myo17b,myo17c,myo17d,myo17e,lyu17}, in which 
antisymmetrized molecular dynamics (AMD) commonly becomes the basis wave function \cite{kanada03}.
The first method is ``tensor-optimized antisymmetrized molecular dynamics'' (TOAMD), 
in which two kinds of the correlation functions with central and tensor operator types
are multiplied to the AMD wave function successively.
In the analysis of $s$-shell nuclei with TOAMD \cite{myo17a,myo17b,myo17c,myo17d}, we can nicely reproduce the results of Green's function Monte Carlo (GFMC) within the double products of the correlation functions.
TOAMD is an extendable framework by increasing the power series of the multiple products of correlation functions, while computational cost becomes large as the mass number increases to $p$-shell nuclei.

The second method is ``high-momentum antisymmetrized molecular dynamics'' (HM-AMD),
in which high-momentum components of nucleons are introduced in the AMD wave function without  correlation functions.
In AMD, nucleon wave function has a Gaussian wave packet with a centroid position in phase space.
We put the large imaginary values in the Gaussian centroids of two nucleons in opposite signs \cite{myo17e,lyu17}, following the idea in Ref. \cite{itagaki17}.
This treatment causes the large-relative-momentum components in nucleon pair in the AMD wave function.
We call the nucleon pair with high-momentum components ``{\it high-momentum pair}''.
In our previous work of HM-AMD \cite{myo17e}, it is confirmed that one high-momentum pair in nuclei gives the equivalent effect to the full space of the two-particle--two-hole (2p--2h) excitations induced by the tensor force.
We obtained this conclusion by comparing HM-AMD with the tensor-optimized shell model \cite{myo05,myo07,myo09}, which treats the 2p--2h excitations fully.

In principle, HM-AMD is a method of superposing the AMD basis states and easy to handle. 
This method is considered to be a promising framework treating the bare $NN$ interaction.
It is an advantage of HM-AMD that this method can be combined with TOAMD.
In the hybrid method of HM-AMD and TOAMD, the AMD wave function becomes the multi-configuration with high-momentum pair and this hybridization was shown to successfully describe the many-body correlations 
of the bare $NN$ interaction in our recent study \cite{lyu17}.
In this hybrid method named ``HM-TOAMD'', one high-momentum pair and the linear terms of the correlation functions are considered simultaneously. 
For the analysis to $p$-shell nuclei, we will extend this new scheme to include the multi-correlation effects by increasing both the number of high-momentum pairs and the power of the correlation functions.

For HM-AMD, it is a fundamental problem to investigate how much this new method can describe the correlations from the $NN$ interaction.
For this aim, in this paper we extend HM-AMD to include up to two kinds of high-momentum pairs in each AMD basis state and superpose them, which extends the variational space of HM-AMD.
We propose a new scheme for constructing the double high-momentum pairs in HM-AMD.
As the development of the previous study of tensor correlation \cite{myo17e}, 
we focus on the description of short-range correlation in HM-AMD.
This study is the first application of the extended HM-AMD; we calculate $^3$H using central potential with short-range repulsion and compare the results with those of TOAMD and GFMC.
We also investigate the properties of the high-momentum pairs in short-range correlation and compare them with those of tensor correlation.
The present extension of HM-AMD becomes the important foundation of HM-TOAMD for the application to finite nuclei with heavier mass systems.

%%%%%%%%%%%%%%%%%%%%%%%%%%%%%%%%%%%%%%%%%%%%%%%%%%%%%%%%%%%%%%%%%%%
\section{High-Momentum Antisymmetrized molecular dynamics (HM-AMD)}\label{sec:AMD}
We first define the AMD wave function $\Phi_{\rm AMD}$ with a Slater determinant of $A$-nucleons,
\begin{eqnarray}
\Phi_{\rm AMD}
&=& \frac1{\sqrt{A!}}\, {\rm det} \left\{\prod_{i=1}^{A} \phi_i(\vc{r}_i) \right\}\,, \qquad
\label{eq:AMD}
\\
\phi(\vc{r}) &=& \left( \frac{2\nu}\pi \right)^{3/4} e^{-\nu(\vc{r}-\vc{Z})^2}\chi_\sigma\chi_\tau.
\label{eq:Gauss}
\end{eqnarray}
The wave function $\phi( \vc{r})$ for nucleon is a Gaussian wave packet with a range parameter $\nu$
and the centroid position $\vc{Z}$.
The spin part $\chi_{\sigma}$ is the up ($\uparrow$) or down ($\downarrow$) component for $z$ direction. 
The isospin part $\chi_{\tau}$ is a proton (p) or neutron (n).
The AMD wave function $\Phi_{\rm AMD}$ is projected on the eigenstates of the angular momentum $J$ and the parity ($\pm$) using the projection operators $P^J_{MK}$ and $P^{\pm}$ as
\begin{eqnarray}
\Psi^{J^\pm}_{MK}
&=& P^J_{MK}P^{\pm} \Phi_{\rm AMD}\, .
\label{eq:projection}
\end{eqnarray}
The angular momentum projection is performed numerically and we take twenty points for each of three Euler angles.

We prepare the AMD wave functions using various sets of the Gaussian centroids $\{\vc{Z}_i\}$
for $i=1,\ldots,A$, and superpose them using the generator coordinate method (GCM).
In AMD+GCM, the total GCM wave function $\Psi_{\rm GCM}$ is the linear combination form of the projected AMD basis states in Eq.~(\ref{eq:projection}) as 
\begin{eqnarray}
   \Psi_{\rm GCM}
&=& \sum_{\alpha} C_{\alpha}  \Psi_\alpha ,
   \label{eq:GCM}
\end{eqnarray}
where the label $\alpha$ is a representative quantum number to identify the projected AMD basis state.
We solve the eigenvalue problem with respect to the Hamiltonian and obtain the total energy $E$ and the expansion coefficients $C_\alpha$ of the GCM wave function.

Next we introduce the high-momentum components in the AMD wave function in HM-AMD.
In this study, we take the case of $^3$H including up to two of nucleon pairs with high-momentum components  as ``high-momentum pair'' in each AMD basis state.
We prepare the basic configuration of $^3$H as the $(0s)^3$ one in AMD with $\vc{Z}= 0 $ for all nucleons.
Following the previous studies \cite{myo17e,lyu17,itagaki17},
we consider the imaginary value of the centroid position $\vc{Z}$ for the nucleon wave packets in Eq.~(\ref{eq:Gauss}), contributing to the nucleon mean momentum as
\begin{eqnarray}
   \frac{\bra \phi | \vc{p} | \phi \ket}{ \bra \phi | \phi \ket }
   &=& 2 \hbar \nu\, {\rm Im}(\vc{Z}),
\end{eqnarray}
where $\vc{p}=-i \hbar \nabla$.
Owing to this relation, the high-momentum component can be included in the AMD wave function with large imaginary value of $\vc{Z}$.
Considering two-nucleon correlations induced by the $NN$ interaction, we focus on the momenta of two nucleons, whose centroid positions are $\vc{Z}_1$ and $\vc{Z}_2$ in the wave packets.
We give the imaginary values in $\vc{Z}_1$ and $\vc{Z}_2$ with the same magnitude in opposite directions as
\begin{eqnarray}
    \vc{Z}_1&=& i \vc{D}, \qquad
    \vc{Z}_2~=~-i \vc{D},
    \label{eq:imaginary}
\end{eqnarray}
where the vector $\vc{D}$ is real and corresponds to the momentum vector of nucleon.
Equation~({\ref{eq:imaginary}) makes the two-nucleon correlations in the AMD wave function, in which two nucleons are excited from the $\vc{D}=0$ state to the finite $\vc{D}$ state.
The relative momentum of two nucleons comes from the relation of $\vc{Z}_1-\vc{Z}_2=2i\vc{D}$,
while the center-of-mass momentum is kept to be zero from the relation of $\vc{Z}_1+\vc{Z}_2=0$.
We call two nucleons with large magnitudes of $\vc{D}$ in Equation~({\ref{eq:imaginary}) ``{\it high-momentum pair}''.
For $^3$H, three kinds of high-momentum pairs are available as
\begin{eqnarray}
1.~~\mbox{p$_\uparrow$ and n$_\uparrow$}   \, ,\qquad
2.~~\mbox{p$_\uparrow$ and n$_\downarrow$} \, ,\qquad
3.~~\mbox{n$_\uparrow$ and n$_\downarrow$} \, .
\label{eq:pair}
\end{eqnarray}
The momentum direction in the high-momentum pair is assigned by using $\vc{D}$.
Following the previous study \cite{myo17e}, we choose three Cartesian coordinates for momentum as
\begin{eqnarray}
    x \mbox{ direction}:\,\vc{D}\,=\,D_x\, \hat{\vc{x}},\qquad
    y \mbox{ direction}:\,\vc{D}\,=\,D_y\, \hat{\vc{y}},\qquad
    z \mbox{ direction}:\,\vc{D}\,=\,D_z\, \hat{\vc{z}},
\end{eqnarray}
where $\hat{\vc{x}}$, $\hat{\vc{y}}$ and $\hat{\vc{z}}$ are the unit vectors of the $x$ $y$, and $z$ directions, respectively.
The lengths $D_x$, $D_y$ and $D_z$ are the shifts of momentum for each direction. 
In the present study, we take the length of $\vc{D}$ from 1 fm to 14 fm in steps of 1 fm for all directions in addition to the $\vc{D}=0$ in the $(0s)^3$ configuration.
Under this condition, we construct the AMD basis states including a single high-momentum pair. This method is called ``single HM-AMD''.

It is noted that the present high-momentum pairs defined in Eq.~(\ref{eq:pair}) in HM-AMD are not always the eigenstates of spin and isospin of two nucleons.
If we superpose the AMD basis states utilizing the exchanges of spin or isospin of two nucleons in each pair, we can construct the two-nucleon basis states with the eigenstates of spin and isospin.
This representation can help to understand the two-nucleon correlations in the spin-isospin channels for many-body system.

In single HM-AMD, following the previous study \cite{myo17e}, it is enough to superpose the basis states with two kinds of momentum directions $\vc{D}$ for the pairs; 
$z$ ($x$) direction parallel (perpendicular) to that of the intrinsic spins, 
where the pairs in $x$ and $y$ directions give the identical effect for $^3$H.

Next, we newly extend HM-AMD so as to include up to two of high-momentum pairs in each AMD basis state.
We explain how to define two pairs using the diagrams in Fig.~\ref{fig:diagram}.
Single high-momentum pair provides the large-relative-momentum component in nucleon pair.
This is expressed schematically in Fig.~\ref{fig:diagram} (a) using the symbol of ``HM$_1$'' and the configuration [12] for the pair, which is introduced in TOAMD in Refs. \cite{myo15,myo17d}.
We follow this expression and utilize the cluster expansion to define the double high-momentum pairs, HM$_1$ and HM$_2$, as is shown in Fig. \ref{fig:diagram} (b).
In this case, double operations of the high-momentum pairs can produce up to four-body correlations in nuclei.
For $^3$H case, we consider up to three-body correlations as a linked (connected) diagram of the configuration [12:13] in Fig. \ref{fig:diagram} (b).

%%%%%%%%%%%%%%%%%%%%%%%%%%%%%%%%%%%%
\begin{figure}[t]
\centering
\includegraphics[width=12.0cm,clip]{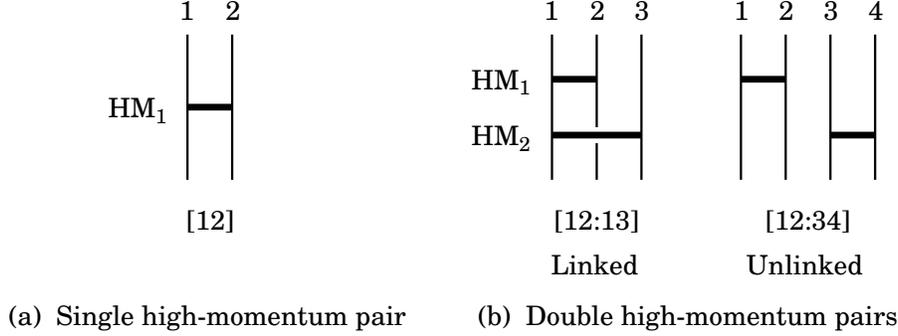}
\caption{Diagrams of the cluster expansion of the single and double high-momentum pairs with HM$_1$ and HM$_2$.
Square brackets below the diagrams represent the configurations of the many-body correlations explained in Refs.~\cite{myo15,myo17d}.}
\label{fig:diagram}
\end{figure}
%%%%%%%%%%%%%%%%%%%%%%%%%%%%%%%%%%%%

We explain explicitly how to construct the basis states for double high-momentum pairs in HM-AMD, named 
``double HM-AMD'' as follows.
\begin{enumerate}
\setlength{\itemsep}{0.25cm}
\setlength{\leftskip}{-0.2cm}
\setlength{\itemindent}{0.0cm}
\item[1)] 
First pair, HM$_1$, is the same as that prepared in the single HM-AMD.
We employ the basis states with two kinds of momentum directions, $z$ and $x$, for the pairs.
\item[2)] 
Second pair, HM$_2$, is connected to HM$_1$ via the first nucleon in Fig.~\ref{fig:diagram} (b), leading to the three-body correlation as [12:13] in the AMD basis states.
We introduce another shift vector $\vc{D}'$ for HM$_2$, and define the centroid parameters of three nucleons as 
\begin{eqnarray}
\vc{Z}_1&=& i\vc{D}\pm i\vc{D}'~,\\
\vc{Z}_2&=&-i\vc{D}~,  \\
\vc{Z}_3&=&        \mp i\vc{D}'~,
\end{eqnarray}
where $\sum_{i=1}^3 \vc{Z}_i=0$ keeping the center-of-mass position zero. For the vector $\vc{D}'$, we take $x$, $y$, and $z$ directions
with the length from 1 fm to 14 fm including the opposite signs.
Here, we impose one condition of $|\vc{Z}_1|=|\vc{D}\pm\vc{D}'|\le 14$ fm
for the linked nucleon in HM$_2$.
We use this setting of HM$_1$ and HM$_2$ cyclically among three nucleons.

\item[3)]
When we treat mass number 4 and beyond, four-body correlation consisting of HM$_1$ and HM$_2$ is incorporated using the unlinked diagram [12:34] in Fig.~\ref{fig:diagram} (b).
\end{enumerate}
We superpose all the AMD basis states with single high-momentum pair and double high-momentum pairs
shown in Fig.~\ref{fig:diagram} considering the above conditions in double HM-AMD.

For the Hamiltonian in the present study, we use the Argonne V4$^\prime$ (AV4$^\prime$) $NN$ central potential with short-range repulsion, 
which is renormalized from the realistic AV18 potential and has four spin-isospin components \cite{wiringa95}.
We expand this potential using Gaussian functions with eight ranges for each component to calculate the interaction matrix elements.

%%%%%%%%%%%%%%%%%%%%%%%%%%%%%%%%%%%%%%%%%%%%
\section{Tensor-optimized antisymmetrized molecular dynamics (TOAMD)}\label{sec:TOAMD}
We briefly explain TOAMD for central correlation, the results of which are compared with those of HM-AMD.
In this study, we consider the central-type correlation function $F_S$ to describe short-range correlation.
We define two kinds of the TOAMD wave functions with single correlation function and double products of the correlation functions as
\begin{eqnarray}
\Phi_{\rm TOAMD}^{\rm single}
&=& (1+F_{S}) \times\Phi_{\rm AMD}~,
\label{eq:TOAMD1}
\\
\Phi_{\rm TOAMD}^{\rm double}
&=& (1+F_{S_1}+F_{S_2}F_{S_3}) \times\Phi_{\rm AMD}~,
\label{eq:TOAMD2}
\\
F_{S}
&=& \sum_{t=0}^{1}\sum_{s=0}^{1}\sum_{i<j}^{A} f^{t,s}_{S}(r_{ij})\,(\boldsymbol\tau_i\cdot \boldsymbol\tau_j)^t\,(\boldsymbol\sigma_i\cdot \boldsymbol\sigma_j)^s \, ,
\label{eq:Fs}
\end{eqnarray}
where $r_{ij}=|\vc{r}_i - \vc{r}_j|$, and $t$ and $s$ are the isospin and spin of two nucleons, respectively.
It is noted that three kinds of $F_{S}$ in double TOAMD in Eq.\,(\ref{eq:TOAMD2}) are independent and variationally determined.
This treatment of the correlation functions extends the variational space of TOAMD in comparison with the Jastrow method \cite{myo17c}.

We use the Gaussian expansion to describe the pair functions $f^{t,s}_{S}(r)$ in Eq.~(\ref{eq:Fs}) as
\begin{eqnarray}
   f^{t,s}_{S}(r)
&=& \sum_{n=1}^{N_G} C^{t,s}_{n}\, e^{-a^{t,s}_{n} r^2}~,
   \label{eq:cr_S}
\end{eqnarray}
where $C^{t,s}_{n}$ and $a^{t,s}_{n}$ are the variational parameters.
We take the basis number $N_G=10$.
The range parameters $a^{t,s}_n$ are optimized in a wide range to express the radial correlation.
Expansion coefficients $C^{t, s}_{n}$ are determined by diagonalizing the Hamiltonian matrix in the 
variation of the total energy in TOAMD \cite{myo17d}.

In the present calculation of TOAMD for $^3$H, we set the $(0s)^3$ configuration for $\Phi_{\rm AMD}$ 
in Eqs.~({\ref{eq:TOAMD1}), ({\ref{eq:TOAMD2}), 
which is obtained in the results of the single TOAMD wave function \cite{myo17d}, and commonly used in HM-AMD.
The value of $\nu$ in the Gaussian wave packet is 0.20 fm$^{-2}$ and 0.095 fm$^{-2}$ for the single and double TOAMD calculations, 
respectively, which are optimized in the energy minimization of $^3$H.

%%%%%%%%%%%%%%%%%%%%%%%%%%%%%%%%%%%%%%%%%%%%
\section{Results}\label{sec:results}

%%%%%%%%%%%%%%%%%%%%%%%%%%%%%%%%%%%%
% AMD/Src4.5.1/Data_AV4p_nu22
% AMD/Src4.5.1/Data_AV4p_nu22_2
\begin{figure}[t]
\centering
\includegraphics[width=7.5cm,clip]{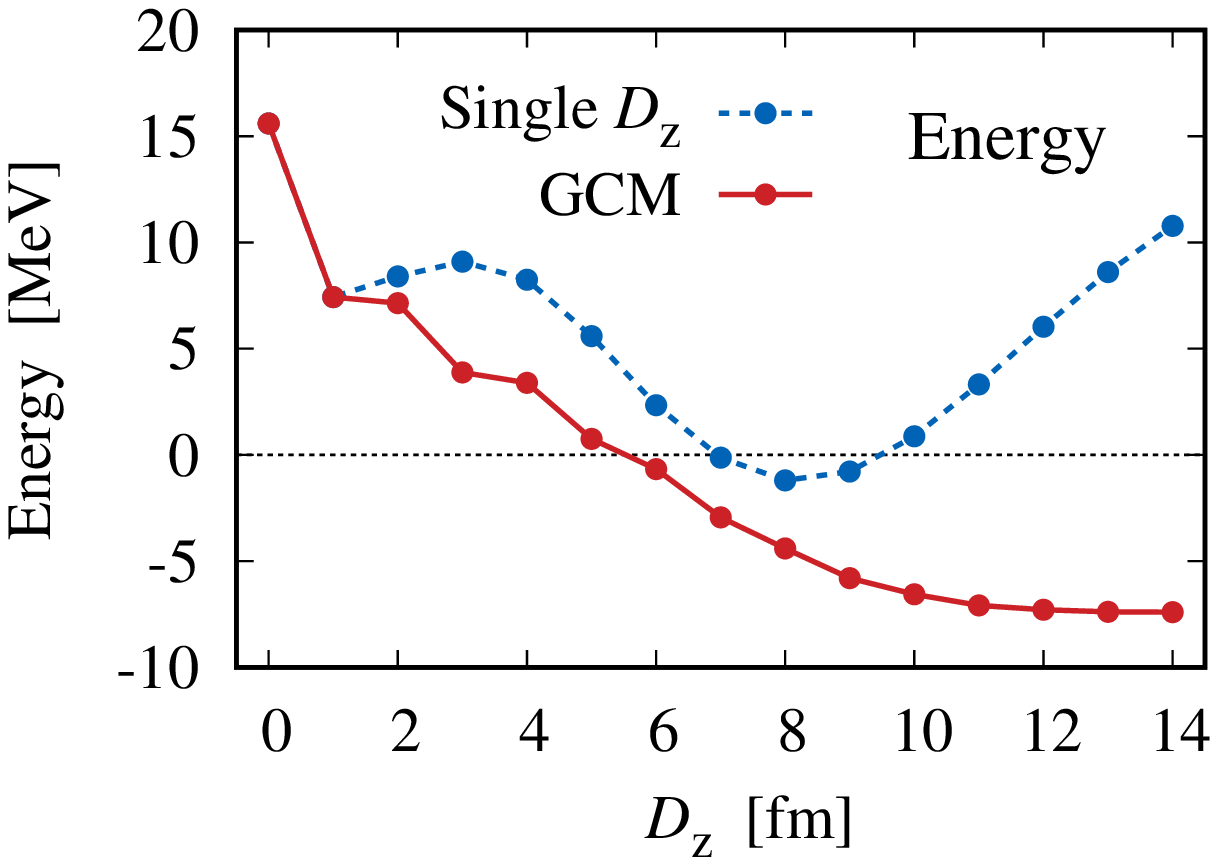}~
\includegraphics[width=7.5cm,clip]{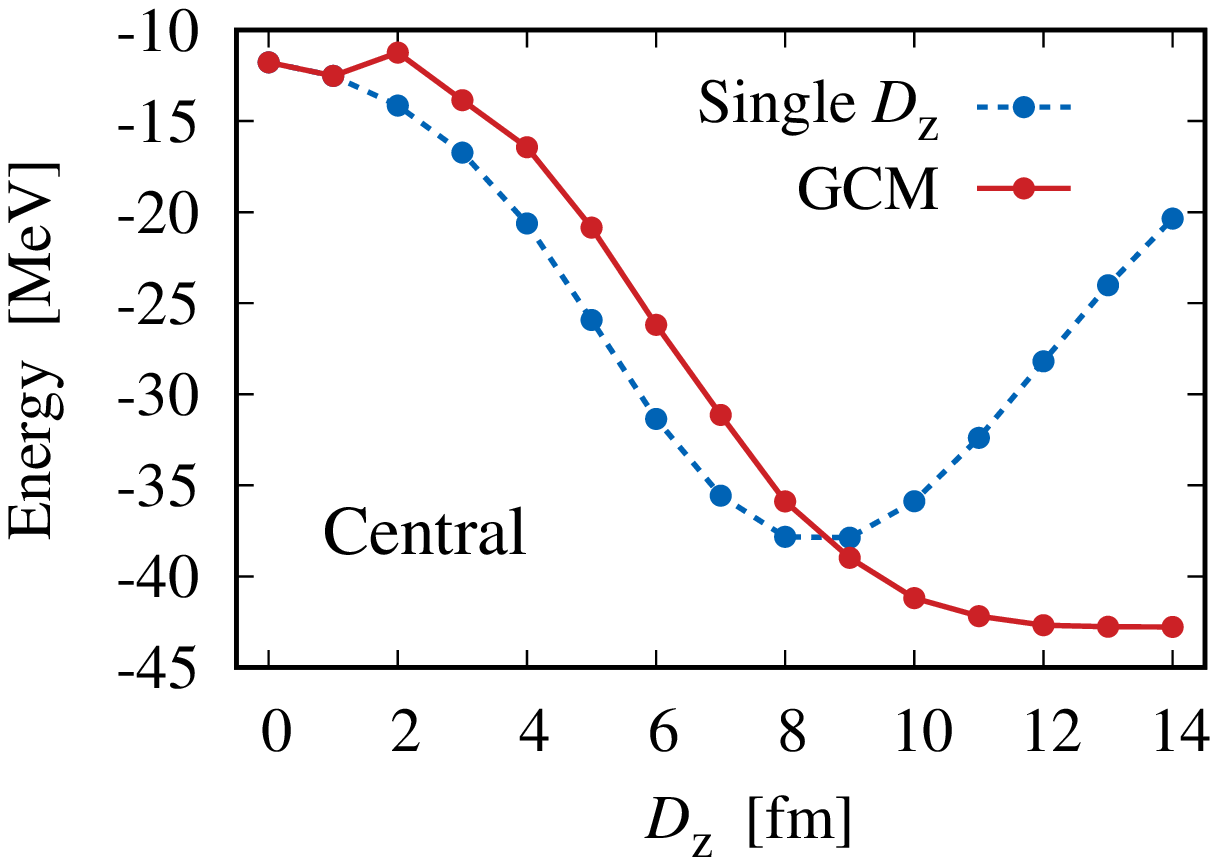}
\caption{Total energy (left) and central matrix element (right) of $^3$H (1/2$^+$) in single HM-AMD with respect to the imaginary shift $D_z$. 
Dotted lines are the results of two basis states with the $(0s)^3$ configuration and a single $D_z$ one.
Solid lines are the results of GCM by adding the basis states successively as increasing $D_z$.}
\label{fig:ene_AMD}
\end{figure}
%%%%%%%%%%%%%%%%%%%%%%%%%%%%%%%%%%%%

%%%%%%%%%%%%%%%%%%%%%%%%%%%%%%%%%%%%
\begin{figure}[t]
\centering
\includegraphics[width=7.5cm,clip]{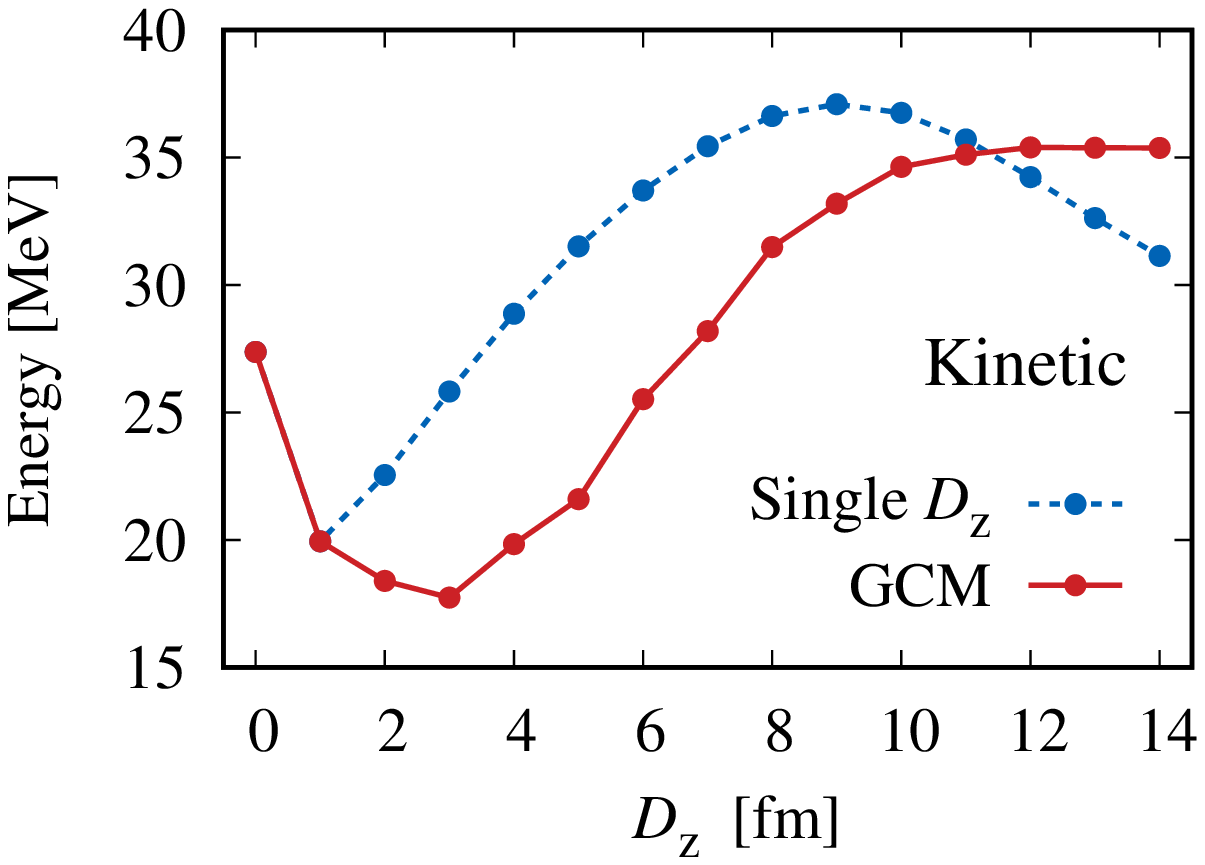}
\includegraphics[width=7.5cm,clip]{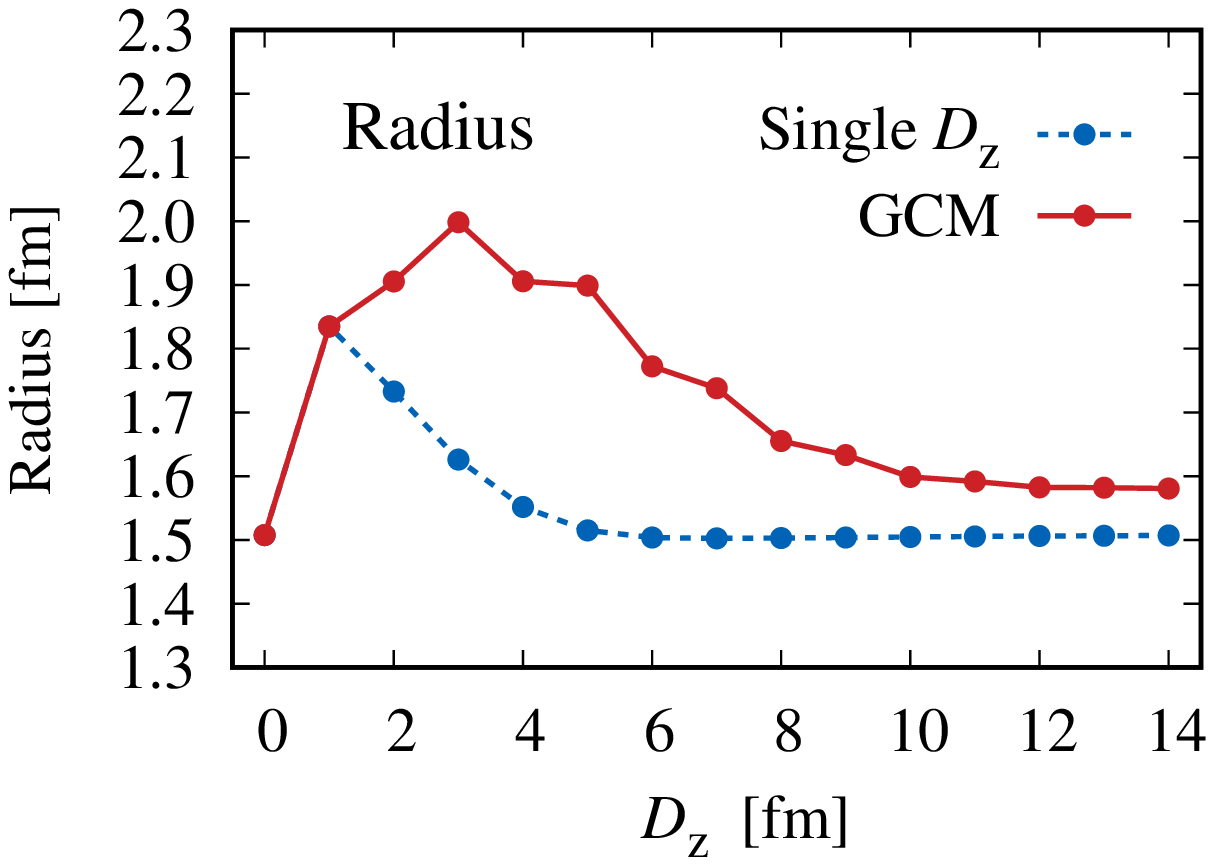}
\caption{Kinetic energy (left) and radius (right) of $^3$H (1/2$^+$) in single HM-AMD with respect to the imaginary shift $D_z$. 
Notations are the same as those in Fig. \ref{fig:ene_AMD}. The unit of radius is fm.
}
\label{fig:kin_AMD}
\end{figure}
%%%%%%%%%%%%%%%%%%%%%%%%%%%%%%%%%%%%

We explain the results of HM-AMD for $^3$H.
We use the range parameter $\nu=0.22$ fm$^{-2}$ for the Gaussian wave packet in Eq. (\ref{eq:Gauss}) in AMD,
which is optimized to minimize the total energy of $^3$H in the double HM-AMD calculation.

First we show the results obtained in the single HM-AMD.
We discuss the behavior of solutions as functions of the shift vector $\vc{D}$ in Eq.~(\ref{eq:imaginary}) in high-momentum pair.
We take the case of $z$ direction with $D_z$, parallel to that of nucleon spins,
and superpose three kinds of nucleon pairs in Eq.~(\ref{eq:pair}). 
In Fig. \ref{fig:ene_AMD}, we plot the energy surface as functions of $D_z$ with dotted lines. 
In this calculation, the $(0s)^3$ configuration with $D_z=0$ and the configuration with a specific $D_z$ are superposed as the two basis cases.
This analysis shows the dependence of the solutions on the shift $D_z$.
On the left-hand side of Fig.~\ref{fig:ene_AMD}, the energy minimum is obtained at $D_z=8$ fm, which provides the mean momentum (wave number) of the single nucleon motion in high-momentum pair as
\begin{eqnarray}
2\nu\cdot {\rm Im}(D_z)&=& 2\cdot 0.22\cdot 8~\simeq~3.5~{\rm fm}^{-1}.
\end{eqnarray}
This value is larger than the Fermi momentum of $1.4$ fm$^{-1}$ by more than twice,
indicating that the high-momentum component is variationally favored in HM-AMD.
It is noted that this value is larger than that in the tensor correlation as 2.5 fm$^{-1}$ \cite{myo17e}.
These results indicate that short-range correlation causes larger momentum component in nuclei than that of tensor correlation,
which is consistent to the analysis of momentum distribution in light nuclei \cite{pieper01_2}.

In Fig. \ref{fig:ene_AMD}, the solid lines show the GCM calculations by adding successively the AMD basis states as increasing $D_z$ from 1 fm to 14 fm. 
The energy convergence is obtained at around $D_z=12$ fm, corresponding to the nucleon mean momentum of 5.3 fm$^{-1}$.

On the right-hand side of Fig. \ref{fig:ene_AMD}, we show the results of the central matrix elements.
The overall behavior of the results is similar to that of the total energy;
the minimum position is commonly obtained at $D_z=8$ fm in the dotted line
and converging matrix elements are obtained at around $D_z=12$ fm.

On the left-hand side of Fig. \ref{fig:kin_AMD}, we show the results of the kinetic energy, similarly to the case of Fig.~\ref{fig:ene_AMD}.
In the dotted line, the largest value is obtained at $D_z=9$ fm, which is close to the energy minimum position as is shown in Fig.~\ref{fig:ene_AMD}.
It is also found that for small values of $D_z$ at around 3 fm, the GCM results show the reduction of the kinetic energy.
This indicates the inclusion of the low-momentum component of nucleon motion in the GCM wave function.
From the results of the converging Hamiltonian components using large values of $D_z$, the present AMD basis states with high-momentum pairs contribute to describe the short-range correlation.

We also show the radius of $^3$H on the right-hand side of Fig. \ref{fig:kin_AMD}.
In general, radius can be sensitive to the low-momentum component of the wave function, and the present results show the convergence of the radius,
which indicates that HM-AMD can treat not only the high-momentum component but also the low-momentum one.
This effect can be taken into account by using the smaller values of $D_z$ in the AMD basis states.

We perform the same analysis for $x$ direction of the high-momentum pair, and confirm that the results are almost identical. 
This means that the short-range correlation is a central-type one and shows the isotropic property with respect to the momentum direction of pair. 
This point is discussed later in more detail.

%%%%%%%%%%%%%%%%%%%%%%%%%%%%%% 
% AMD/Src4.5.1/Data_AV4p_nu22
% AMD/Src4.5.1/Data_AV4p_nu22_2
\begin{table}[t]
\begin{center}
\caption{Energies and Hamiltonian components of $^3$H ($1/2^+$) in HM-AMD for three kinds of nucleon pairs in units of MeV. The momentum direction is $z$ direction.}
\label{tab:AMD_GCM} 
	\begin{tabular}{lrrrrrrrrrrrrrr}
\noalign{\hrule height 0.5pt}
       & \lw{$(0s)^3$}  & \multicolumn{4}{c}{$z$}              \\
        \cline{3-6}
              &         & $p_\uparrow$-$n_\uparrow$ &$p_\uparrow$-$n_\downarrow$ & $n_\uparrow$-$n_\downarrow$ & all  \\
\noalign{\hrule height 0.5pt}
Total  energy &$ 15.60$ &$  7.20$&$  5.34$ &$  5.24$&$ -7.40$  \\
Kinetic energy&$ 27.37$ &$ 26.15$&$ 25.54$ &$ 24.66$&$ 35.38$  \\
Central force &$-11.70$ &$-18.95$&$-20.20$ &$-19.42$&$-42.78$  \\
\noalign{\hrule height 0.5pt}
\end{tabular}
\end{center}
\end{table}
%%%%%%%%%%%%%%%%%%%%%%%%%%%%%%

%%%%%%%%%%%%%%%%%%%%%%%%%%%%%% 
% AMD/Src4.5.1/Data_AV4p_nu22
% AMD/Src4.5.1/Data_AV4p_nu22_2
\begin{table}[t]
\begin{center}
\caption{Same analysis as done for Table \ref{tab:AMD_GCM}. The momentum direction is $x$ direction.}
\label{tab:AMD_GCM2} 
	\begin{tabular}{lrrrrrrrrrrrrrr}
\noalign{\hrule height 0.5pt}
              & \multicolumn{4}{c}{$x$}\\
        \cline{2-5}
              & $p_\uparrow$-$n_\uparrow$ &$p_\uparrow$-$n_\downarrow$ & $n_\uparrow$-$n_\downarrow$ & all \\
\noalign{\hrule height 0.5pt}
Total  energy &$  6.64$&$  4.24$&$  5.21$&$ -7.59$ \\
Kinetic energy&$ 26.88$&$ 26.02$&$ 24.69$&$ 35.56$ \\
Central force &$-22.49$&$-21.78$&$-19.49$&$-43.15$ \\
\noalign{\hrule height 0.5pt}
\end{tabular}
\end{center}
\end{table}
%%%%%%%%%%%%%%%%%%%%%%%%%%%%%%

%%%%%%%%%%%%%%%%%%%%%%%%%%%%%% 
\begin{table}[t]
\begin{center}
\caption{Same analysis as done for Tables \ref{tab:AMD_GCM} and \ref{tab:AMD_GCM2}, superposing the basis states with momenta both for $z$ and $x$ directions. For $n_\uparrow$-$n_\downarrow$ part, the results does not change from those with $x$ direction as shown in Table \ref{tab:AMD_GCM2}.}
\label{tab:AMD_GCM3} 
	\begin{tabular}{crrrrrrr}
\noalign{\hrule height 0.5pt}
       & \multicolumn{4}{c}{$z$ and $x$}  \\               
       \cline{2-5}\cline{6-6}
              & $p_\uparrow$-$n_\uparrow$ & $p_\uparrow$-$n_\downarrow$ & $n_\uparrow$-$n_\downarrow$ & all  \\
\noalign{\hrule height 0.5pt}
Total  energy & $  5.61$&$  3.80$&$  5.21$&$ -7.62$ \\
Kinetic energy& $ 27.25$&$ 26.28$&$ 24.69$&$ 35.56$ \\
Central force & $-21.66$&$-22.48$&$-19.49$&$-43.18$ \\
\noalign{\hrule height 0.5pt}
\end{tabular}
\end{center}
\end{table}
%%%%%%%%%%%%%%%%%%%%%%%%%%%%%%

We discuss the roles of three kinds of pairs in Eq.~(\ref{eq:pair}) for $^3$H. 
In Tables \ref{tab:AMD_GCM} and \ref{tab:AMD_GCM2}, we show the results of each pair with $z$- and $x$-directions, 
where the basis states with various $D_z$ and $D_x$ are superposed.
It is found that three kinds of pairs provide the similar results for all directions;
energy gain is about 10 MeV from the $(0s)^3$ configuration.
Among the pairs, the $p_\uparrow$-$n_\uparrow$ pair shows the rather large positive energies. 
This is understood from the state dependence of the correlations of the pair.
The $p_\uparrow$-$n_\uparrow$ pair can describe the correlation only for spin-triplet component,
on the other hand, the $p_\uparrow$-$n_\downarrow$ and $n_\uparrow$-$n_\downarrow$ pairs can describe the correlations of the spin-singlet component as well as the triplet one.
This property results in the larger energy gain of the $p_\uparrow$-$n_\downarrow$ and $n_\uparrow$-$n_\downarrow$ pairs than that of the $p_\uparrow$-$n_\uparrow$ pair.

In each direction, ``all''  is the result superposing the basis states with three kinds of pairs. 
It is found that $z$ and $x$ directions of high-momentum pairs provide the similar results.
This means the isotropic property of the short-range correlation, which is different from the case of the tensor correlation \cite{myo17e,feldmeier11},
in which $z$ direction is favored than $x$ direction for the high-momentum pair and 
this is understood from the property of the tensor operator $S_{12}$.

It is also found that each pair gives the unbound state of $^3$H and the superposition of all pairs makes the bound state of $^3$H.
This property is different from the tensor force case, 
in which the contribution of proton-neutron pair is dominant \cite{myo17e}.
In the present case, the superposition of proton-neutron pairs with $p_\uparrow$-$n_\uparrow$ and $p_\uparrow$-$n_\downarrow$,
provides the total energy of $1.27$ MeV for $z$ direction and $1.14$ MeV for $x$ direction, still unbound.
These results mean that all kinds of pairs consisting of protons and neutrons are necessary to be treated for short-range correlation,
because the short-range repulsion exists between all kinds of nucleon pairs.

%%%%%%%%%%%%%%%%%%%%%%%%%%%%%%%%%%%%%%%%%%%%%
In Table \ref{tab:AMD_GCM3}, we superpose the high-momentum pairs with $z$ and $x$ directions and 
``all'' is the final result using the available configurations for high-momentum pairs in single HM-AMD. It is noted that each AMD basis state involves one high-momentum pair.
The resulting energy of $^3$H in single HM-AMD is $-7.62$ MeV, which is underbound by about 1.3 MeV with respect to the GFMC result of $-8.99(1)$ MeV \cite{pieper01_2}.
This difference requires the second high-momentum pair, HM$_2$, as is shown in Fig.~\ref{fig:diagram} (b) in double HM-AMD.
For comparison between HM-AMD and TOAMD, the single TOAMD in Eq.~(\ref{eq:TOAMD1}) gives the energy of $-7.70$ MeV, which is very close to the value of the present single HM-AMD.

%%%%%%%%%%%%%%%%%%%%%%%%%%%%%%%%%%%%
\begin{figure}[t]
\centering
\includegraphics[width=7.5cm,clip]{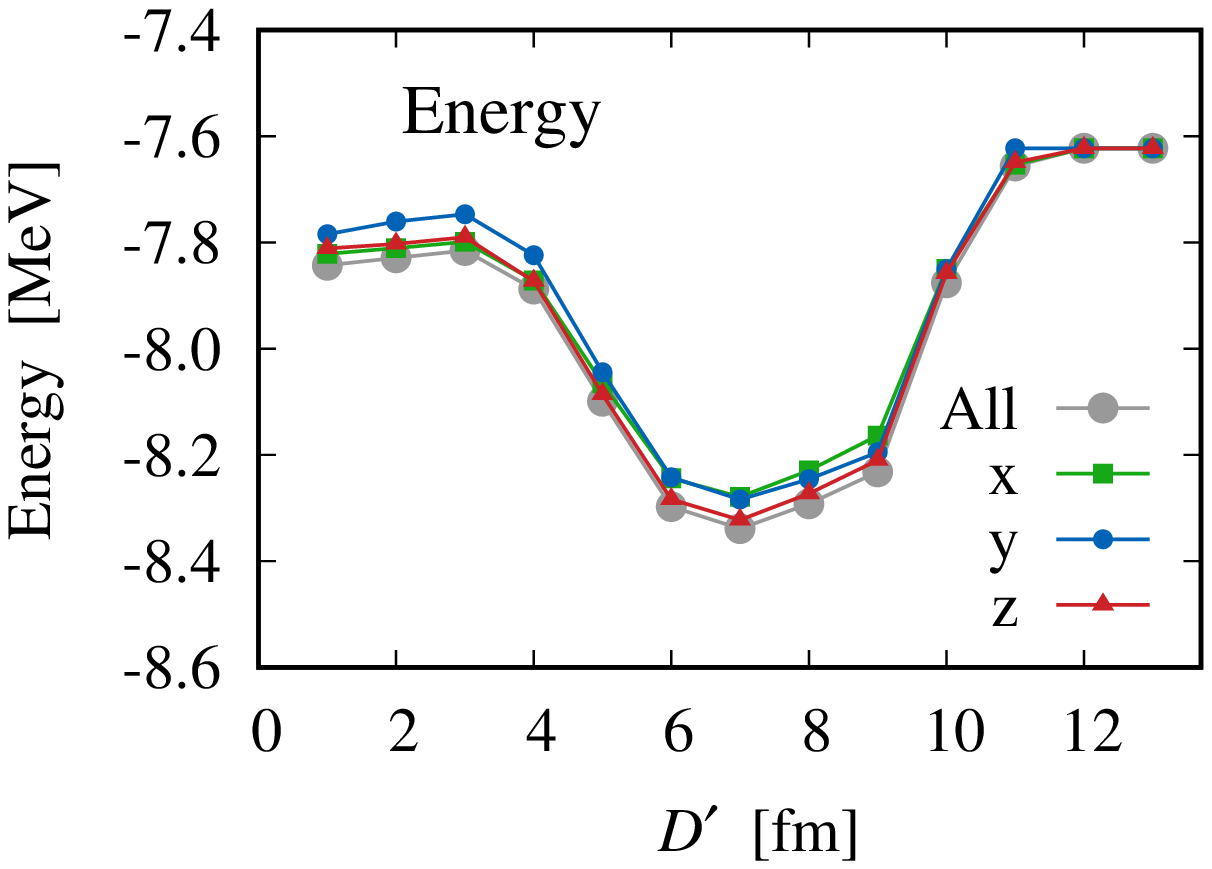}
\includegraphics[width=7.5cm,clip]{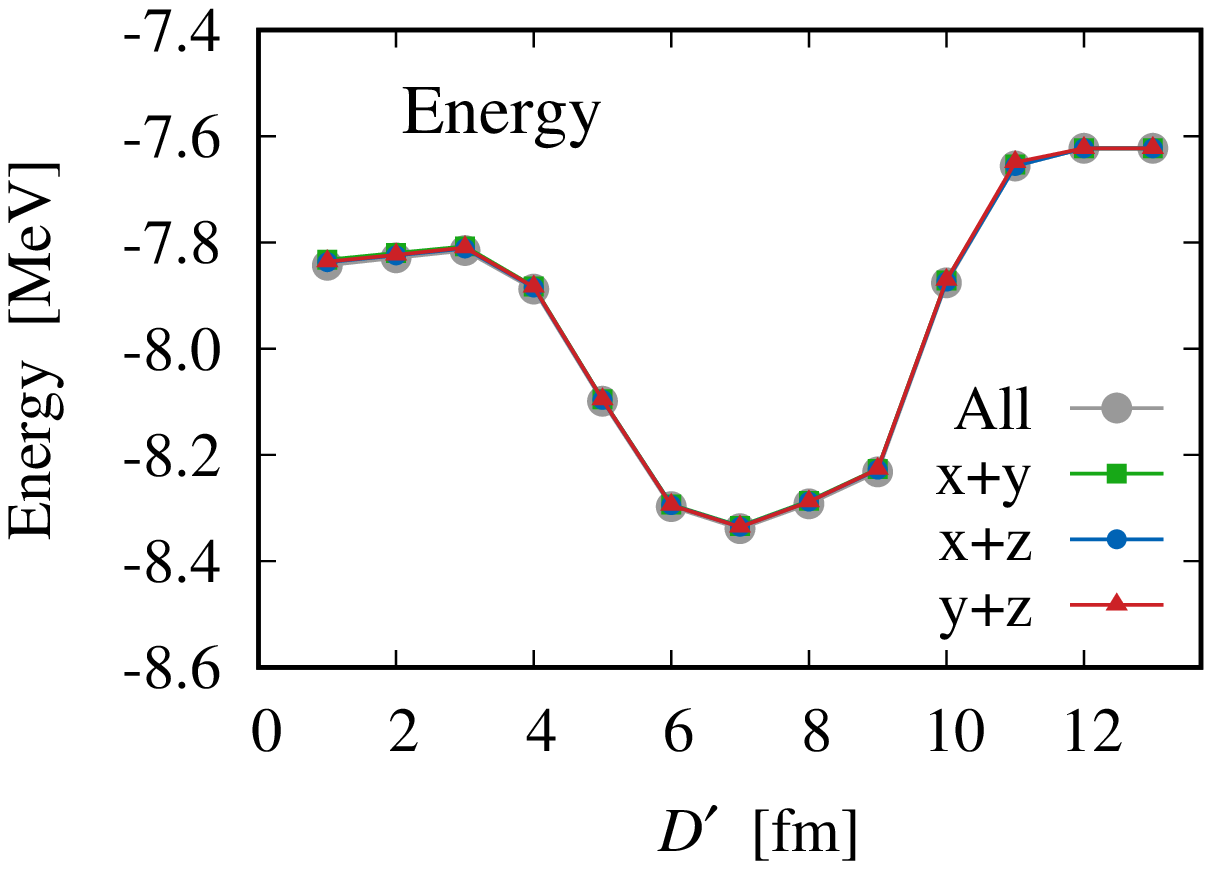}
\caption{Total energy of $^3$H (1/2$^+$) in double HM-AMD with respect to the magnitude of the imaginary shift $D'=|\vc{D}'|$ in the second pair.
Left: one momentum-direction ($x$, $y$, and $z$) is adopted for second pairs. 
Right: two momentum-directions are adopted for second pairs.  
``All'' indicates the results including all three directions in the second pairs for comparison. 
}
\label{fig:double}
\end{figure}
%%%%%%%%%%%%%%%%%%%%%%%%%%%%%%%%%%%%

Finally, we show the results of double HM-AMD using up to two kinds of high-momentum pairs,
in which the additional basis states of the second high-momentum pairs (HM$_2$) are included for all three directions and all three kinds of pairs.
Before showing the final results, we investigate the properties of HM$_2$ in the following way; we limit HM$_2$ to a specific momentum shift $D^\prime$ for all three directions commonly
as $D^\prime=|D^\prime_x|=|D^\prime_y|=|D^\prime_z|$, and search for the energy minimum with respect to $D^\prime$ in double HM-AMD.
This calculation is similar to the case of single HM-AMD as shown in Fig.~\ref{fig:ene_AMD}.
In addition, we investigate the individual effects of the momentum directions of $x$, $y$, and $z$ in HM$_2$.

On the left-hand side of Fig.~\ref{fig:double}, we show the results of double HM-AMD, in which all basis states of the first high-momentum pairs (HM$_1$) are included 
and one momentum-direction is adopted for HM$_2$ with the shift $D'$. ``All'' indicates the results including all three directions in HM$_2$ for comparison.
For each of three directions, energy minima are commonly obtained at $D^\prime=7$ fm,
which is close to $D_z=8$ fm in the results of HM$_1$, as is shown in Fig.~\ref{fig:ene_AMD}. 
This means that the first and second high-momentum pairs cause almost the same momentum component from variational point of view.

It is found that three kinds of momentum-directions give similar energies at each value of $D'$ and their differences are around 40 keV at the energy minimum point.
Among three directions, $z$ direction, parallel to the spin direction, is rather favored energetically,
and provides the energy close to the calculation including all directions by less than 20 keV at the energy minimum point.
This indicates the one momentum-direction effectively describes the contribution of HM$_2$ in the double HM-AMD. 

On the right-hand side of Fig.~\ref{fig:double}, we show the results, in which two momentum-directions are adopted for HM$_2$. 
It is found that three kinds of combinations of two directions provide the energies very close to each other for all ranges of $D'$.
The energy difference from the value of all three directions are few keV at the energy minimum point.
This fact indicates that two momentum-directions of HM$_2$ can provide the converging solutions in the double HM-AMD.
The present analysis of the momentum direction in high-momentum pairs is useful to prepare the basis states of HM-AMD for applying to heavier mass system.

%%%%%%%%%%%%%%%%%%%%%%%%%%%%%% 
\begin{table}[t]
\begin{center}
\caption{Total energies and Hamiltonian components of $^3$H ($1/2^+$) in HM-AMD with single and double high-momentum pairs in units of MeV.}
\label{tab:compare}
\begin{tabular}{lrrrrrr}
\noalign{\hrule height 0.5pt}
        & \multicolumn{2}{c}{HM-AMD}    && \multicolumn{2}{c}{TOAMD} & \lw{GFMC \cite{pieper01_2}}   \\
        \cline{2-3}\cline{5-6}
        & Single            &  Double   && Single       &  Double    &             \\
\noalign{\hrule height 0.5pt}
Total energy   & $-7.62 $   & $ -8.90$  && $ -7.70$     & $-8.97$   &$-8.99(1)$ \\
Kinetic energy & $ 35.56$   & $ 36.89$  && $ 34.81$     & $37.20$   &$     $    \\
Central force  & $-43.18$   & $-45.79$  && $-42.51$     &$-46.17$   &$     $    \\ \hline
\noalign{\hrule height 0.5pt}
\end{tabular}
\end{center}
\end{table}
%%%%%%%%%%%%%%%%%%%%%%%%%%%%%%

We show the results of double HM-AMD superposing the basis states of all three directions of HM$_2$ with various sizes of the shift vector $\vc{D}^\prime$.
In Table \ref{tab:compare}, we list the results of the total energies and Hamiltonian components of $^3$He in comparison with those of TOAMD and GFMC.
For HM-AMD, addition of second high-momentum pair gains the total energy by about 1.3 MeV.
As a result, we confirm a very nice agreement among double HM-AMD, double TOAMD, and GFMC.
It is noted that the double TOAMD provides the very close results to GFMC.
These results indicate that HM-AMD with up to two high-momentum pairs sufficiently describes the short-range correlations in the central force.
In addition, the high-momentum pairs along the $x$, $y$, and $z$ directions nicely work to treat the correlations from the $NN$ interaction.

%%%%%%%%%%%%%%%%%%%%%%%%%%%%%%%%%%%%%
\section{Summary}\label{sec:summary}
We developed a new variational method of ``high-momentum antisymmetrized molecular dynamics'' (HM-AMD), 
which includes up to two kinds of nucleon pairs with high-momentum components for short-range correlation.
This method is based on the superposition of the AMD wave functions.
In HM-AMD, we express the high-momentum components of nucleon motion caused by the short-range correlation in nuclei
in terms of the imaginary positions of the Gaussian centroids in opposite signs for nucleon pair, which is called ``{\it high-momentum pair}''.
In this study, we include up to two kinds of high-momentum pairs in each AMD basis state by utilizing the cluster expansion. 
This is an extension of the previous study of HM-AMD including single high-momentum pair \cite{myo17e}.
In HM-AMD, the AMD basis states are superposed with various momentum components and directions by using high-momentum pairs.

We show the reliability of the extended HM-AMD by showing the results of $^3$H using AV4$^\prime$ central interaction; we obtain the total energy, which is close to that of the {\it ab initio} calculation. 
This means that HM-AMD can express the short-range correlation.
In addition, the solutions of HM-AMD agree well with those of ``tensor-optimized AMD'' (TOAMD) for each Hamiltonian component. 
This fact means that HM-AMD and TOAMD are regarded as the identical framework to treat the correlation induced by the $NN$ interaction.
It is also shown that the isotropic property of short-range correlation for the direction of the high-momentum pairs, which is different from that of tensor correlation.
All kinds of high-momentum pairs of protons and neutrons should be considered for short-range correlation, while tensor correlation is dominated in the proton-neutron pairs \cite{myo17e}.

The present analysis focus on the short-range correlation in the central force.
It is interesting to apply this framework to nuclei using a bare $NN$ interaction with tensor force as well as the short-range repulsion, which is in progress \cite{isaka17}.
In this case, the short-range and tensor correlations should be described simultaneously.
It is interesting to investigate two different characters of the correlations in HM-AMD.

In the present HM-AMD, we express two-body correlations by using high-momentum pairs, and propose the method to introduce many-body correlations from the multiple products of the two-body correlations.
It is interesting to investigate the roles of each higher-body correlation such as four-body correlation in addition to three-body correlation in heavier mass system. 

The present HM-AMD is able to be combined with TOAMD using the correlation functions of the central and tensor operator types.
In this case, effects of correlations can be taken into account by using not only the correlation functions in TOAMD, but also the high-momentum pairs in TOAMD.
We call this new hybrid scheme ``HM-TOAMD'' \cite{lyu17}. 
In Ref. \cite{lyu17}, it is confirmed that the HM-TOAMD with single high-momentum pair and single correlation functions provides the equivalent solutions to those of TOAMD within the double products of the correlation functions.
This hybrid method is a promising and efficient framework for the application to $p$-shell nuclei, where we shall increase the multiple correlations for the description of many-body nuclear systems.

%\section*{Acknowledgments}
\ack
This work was supported by the JSPS KAKENHI Grants No. JP15K05091, and the JSPS Japan-France Joint Research Project and IN2P3 project “Neutron-rich unstable light nuclei”.
We thank Dr. Mengjiao Lyu, Dr. Masahiro Isaka, Professor Hiroshi Toki, Professor Kiyomi Ikeda, Professor Hisashi Horiuchi, Dr. Tadahiro Suhara, and Professor Taiichi Yamada for useful discussions.

%\vfill\pagebreak
%%%%%%%%%%%%%%%%%%%%%%%%%%%%%%%%%%%%%%%%%%%%%%%%%%%%%%%%%%%%%
%\section*{References}
\nc\PTEP[1]{Prog.\ Theor.\ Exp.\ Phys.,\ \andvol{#1}} %%Added by TD 12/06/14
\nc\PPNP[1]{Prog.\ Part.\ Nucl.\ Phys.,\ \andvol{#1}} %%Added by TD 12/06/14

\end{document}